\documentclass[
reprint,
amsmath,
amssymb,
aps,
prc
]{revtex4-2}

\usepackage{graphicx}
\usepackage{dcolumn}
\usepackage{bm}
\usepackage{subcaption}
\usepackage{caption}
\usepackage{float}
\usepackage{hyperref}
\usepackage{lineno}
\usepackage{xcolor}
\hypersetup{colorlinks = true,linkcolor={blue},citecolor={red},urlcolor={blue}}  
\usepackage{tikz,xcolor}
\definecolor{lime}{HTML}{A6CE39}
\DeclareRobustCommand{\orcidicon}{
	\begin{tikzpicture}
	\draw[lime, fill=lime] (0,0) 
	circle [radius=0.16] 
	node[white] {{\fontfamily{qag}\selectfont \tiny ID}};
	\draw[white, fill=white] (-0.0625,0.095) 
	circle [radius=0.007];
	\end{tikzpicture}
	\hspace{-2mm}
}
\foreach \x in {A, ..., Z}{%
	\expandafter\xdef\csname orcid\x\endcsname{\noexpand\href{https://orcid.org/\csname orcidauthor\x\endcsname}{\noexpand\orcidicon}}
}

\begin{document}
\title{On the light front surfaces in the phase space of hadrons in heavy-ion collisions}
\author{Rahul Ramachandran Nair \footnote{Orcid ID: 0000-0001-8326-9846}}
\email{physicsmailofrahulnair@gmail.com}
\affiliation{National Centre For Nuclear Research, 02-093 Warsaw, Poland}
\altaffiliation{The work was performed while the author was affiliated with the National Centre For Nuclear Research, Warsaw, Poland. Present address: Sreeraj Bhavan, Karukachal PO, Kottayam(Dist), Kerala, India- 686540} 
\date{\today}
\begin{abstract}
Surfaces corresponding to the constant values of the light front variable forms paraboloids in the phase space of the inclusively produced hadrons in heavy-ion collisions. In this paper, it is demonstrated with the simulated Au-Au collisions at $\sqrt{s} = 200$ GeV using the UrQMD event generator that the paraboloidal surface defined by a certain constant value of the light front variable divides the particles in the phase space into two groups such that the distributions of the square of the transverse momentum, the polar angle and the light front variable of particles belonging to one among them can be described with the Boltzmann distribution parameterised by the mass-dependent temperature.
\end{abstract}
\maketitle
\section{Introduction}\label{sec:1}
\begin{figure}[ht!]
\includegraphics[width=0.45\textwidth]{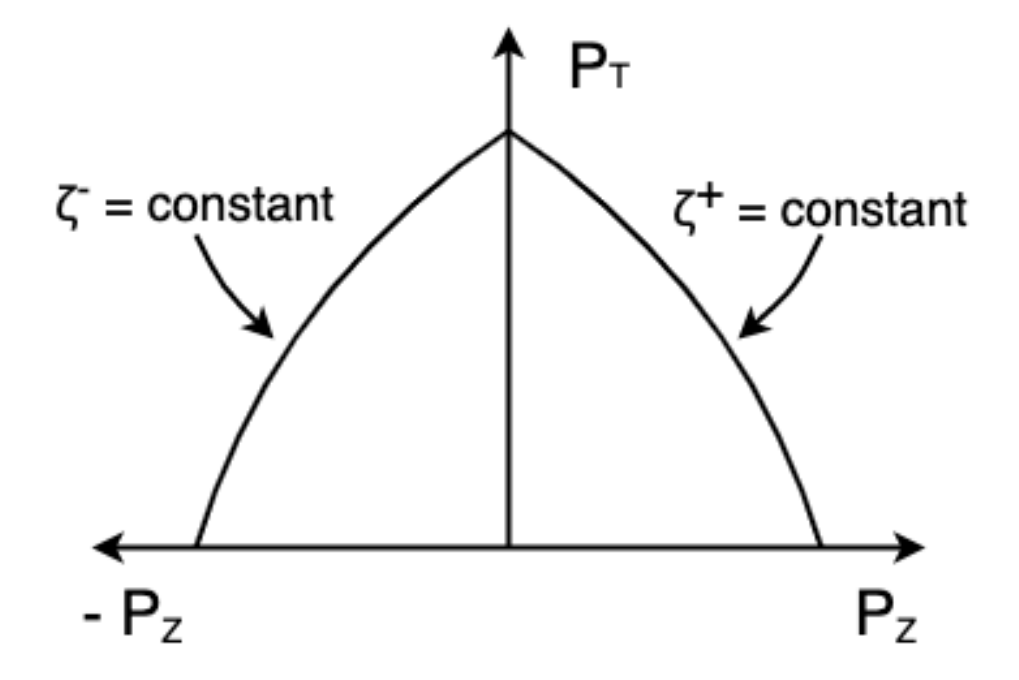} 
\caption{Surfaces of constant $\zeta^{\pm}$ in the Peyrou plot.}
\label{ConstSurf}
\end{figure}

The light front variables were introduced by Dirac  and are used across the spectrum in nuclear and particle physics \cite{Dirac49}. A scale-invariant form of light front variable was proposed in \cite{Garsevanishvili78} for studying the inclusive distribution of particles in hadron-hadron and nucleus-nucleus interactions. The properties of these variables and the analysis based on them were described and performed in \cite{Garsevanishvili79,Amaglobeli99,Djobava03, Chkhaidze2006}. If the four-momentum $p_{\mu} = (p_{0}, \boldsymbol{p})$ of the particle is on the mass shell, then the light front variable together with the transverse momentum defines a horospherical coordinate system on the corresponding mass shell hyperboloid given by $p_{0}^{2} - \boldsymbol{p}^{2} = m^{2}$. It can be shown that the corresponding hyperboloid in the velocity space is a realization of the Lobachevsky space with constant negative curvature \cite{Djobava03,VILENKIN, Garsevanishvili1971}. The proposed light front variable in the centre of mass frame has the following form:
\begin{equation} \label{EqnXi}
\xi^{\pm} = \pm \frac{E \pm p_{z}}{\sqrt{s}} = \pm \frac{E + |p_{z}|}{\sqrt{s}}  \end{equation}
where, $s$ is the Mandelstam variable, $p_{z}$ is the longitudinal component of the momentum and E is the energy of the particle. The positive sign in Eq.\eqref{EqnXi} is used for the right-hand side hemisphere and the negative sign is used for the left-hand side hemisphere. The region $\xi^{\pm} < m/\sqrt{s}$ in the $\xi^{\pm}$ spectra of particles integrated over all the values of the square of the transverse momentum $p_{T}^{2}$ is kinematically forbidden\cite{Djobava03}. It was observed in the earlier studies using the low energy hadron-hadron and nucleus-nucleus collisions, that there exist a peak in the $\xi^{\pm}$ distribution of inclusively produced $\pi^{\pm}$ mesons around a small value of $\xi^{\pm}$ denoted as $\tilde{\xi}^{\pm}$ \cite{Garsevanishvili79,Amaglobeli99,Djobava03}. The $cos(\theta)$ distribution of $\pi^{\pm}$ mesons with $\xi^{\pm} < \tilde{\xi}^{\pm}$ was found to be flat compared to the same distribution of those $\pi^{\pm}$ mesons with $\xi^{\pm} > \tilde{\xi}^{\pm}$ in those collisions.  Similarly, the  $p_{T}^{2}$ distributions of the $\pi^{\pm}$ mesons in these two groups were observed to have different slopes. It lead to the conclusion that the $\pi^{\pm}$ mesons with $\xi^{\pm} < \tilde{\xi}^{\pm}$ produced in those collisions are a thermalised group of particles. To enlarge the scale in the region of smaller $\xi$ values, a convenient variable $\zeta$ can be defined as
\begin{equation}\label{zetadef}
\zeta^{\pm} = \mp \ln(|\xi^{\pm}|)
\end{equation}
Surfaces of constant $\zeta^{\pm}$ form paraboloids in the Peyrou ($p_{T} - p_{z}$) plot of the inclusively produced particles in heavy-ion collisions as shown in Fig.\ref{ConstSurf}. 
The invariant differential cross-section in terms of the light front variables can be written as
\begin{equation}
E\frac{d\sigma}{d\textbf{p}} = \frac{|\xi^{\pm}|}{\pi}\frac{d^2\sigma}{d\xi^{\pm}dp_T^2} = \frac{1}{\pi}\frac{d^2\sigma}{d\zeta^{\pm}dp_T^2}
\end{equation}
To check the hypothesis of thermalisation, $\zeta^{\pm}$ distributions were made for the $\pi^{\pm}$ mesons in the inclusive reaction $ p\bar{p}\longrightarrow \pi^{\pm} + X$ at 22.4 GeV \cite{Amaglobeli99}. To describe the spectra of the $\pi^{\pm}$ mesons  with  $\xi^{\pm} < \tilde{\xi}^{\pm} $ (or equivalently $\zeta^{\pm} > \tilde{\zeta}^{\pm}$), energy distribution of the following form 
\begin{equation}\label{eqboltz}
f(E) \sim  exp(-E/T)
\end{equation} 
was proposed which corresponds to a system that has reached thermal equilibrium. It was found that the $\zeta^{\pm}$ distribution of $\pi^{\pm}$ mesons with $\zeta^{\pm} > \tilde{\zeta}^{\pm}$ can be described using the integral relation 
\begin{equation}
\frac{1}{\pi}\frac{dN}{d\zeta} \sim \int_0^{p_{T,max}^2} E f(E)dp_T^2
\label{ZetaInt}
\end{equation}
where $p_{T,max}^2$ in Eq.\eqref{ZetaInt} is given by 
\begin{equation}
p_{T,max}^2 = (\xi\sqrt{s})^{2} - m^{2} 
\end{equation}
Corresponding $cos(\theta)$ distribution of $\pi^{\pm}$ mesons with $\zeta^{\pm} > \tilde{\zeta}^{\pm}$ was found to be described by the following expression:
\begin{equation}
\frac{dN}{dcos(\theta)} \sim \int_0^{p_{max}} f(E) p^2dp
\label{CosInt}
\end{equation}
where $p_{max}$ is given by
\begin{equation}
p_{max}  = \frac{-\tilde\xi\sqrt{s}cos(\theta) + \sqrt{(\tilde\xi\sqrt{s})^2  - m^2 sin^2(\theta)}}{sin^2(\theta)}
\label{pmax}
\end{equation}
Finally, the $p_{T}^{2}$ distribution of $\pi^{\pm}$ mesons with $\zeta^{\pm} > \tilde{\zeta}^{\pm}$ was found to be describable with the following relation:
\begin{equation}
\frac{dN}{dp_T^2} \sim \int_0^{p_{z,max}} f(E)dp_Z
\label{PtSqInt}
\end{equation}
where $p_{z,max}$ is given by the expression
\begin{equation}
p_{z,max}  = \frac{m^2 + p_T^2 - (\tilde\xi\sqrt{s})^2}{-2\tilde{\xi}\sqrt{s}}
\label{pzmax}
\end{equation}
The limits of integration in Eq.\eqref{ZetaInt}, Eq.\eqref{CosInt} and Eq.\eqref{PtSqInt} given by  $p_{T,max}^2$, $p_{max}$ and $p_{z,max}$ respectively are determined by the boundaries of the paraboloid defined by $\tilde{\xi}^{\pm}$. This simple model could describe the respective spectra of the charged pions, leading to the conclusion that a thermalisation has been reached in the system. A similar analysis was performed for the charged pions in nucleus-nucleus collisions in \cite{Djobava03} and the conclusions remained intact. The analysis of the kind mentioned above is termed as the 'light front analysis'. There were also criticisms raised against the light front scheme of analysis as well \cite{Levchenko}. As per the arguments in \cite{Levchenko}, the observations made in \cite{Garsevanishvili78, Garsevanishvili79} about the light front 'critical' surface in the phase space of hadrons at low energy, allowing to single out the isotropic pions was 'fortuitous' because of the low mass of $\pi^{\pm}$ mesons and lower values of the available centre of mass-energy in the collisions analysed in those studies. It was also concluded that the observed isotropisation should disappear for particles with higher masses. To justify these arguments, an estimation of the angular distribution for the heavier particles were made using a Monte Carlo event generator described in \cite{Levchenko}. However, the light front analysis was not attempted in \cite{Levchenko} with the fits using Eq.\eqref{ZetaInt}, Eq.\eqref{CosInt} and Eq.\eqref{PtSqInt} taking the energy distribution as in Eq.\eqref{eqboltz}. In this paper, a description of the light front analysis of inclusively produced hadrons in the Au-Au collisions at $\sqrt{s} =200$ GeV simulated with an event generator that implements the Ultrarelativistic Quantum Molecular Dynamical (UrQMD) model of heavy-ion collisions is presented \cite{BASS1998255, Bleicher_1999}. It is shown that in contradiction to the arguments made in \cite{Levchenko}, the light front variable based scheme of analysis works at RHIC energies for the particles of mass heavier than pions within the context of the UrQMD framework.
\section{Scheme of analysis}\label{sec:2}
In the analysis presented here, we define two regions in the phase space of the particles as follows:\\
$\bullet$ Region-1:  $\xi^{\pm} < \tilde{\xi}^{\pm} $ or   $\zeta^{\pm} > \tilde{\zeta}^{\pm} $\\
$\bullet$ Region-2:  $\xi^{\pm} > \tilde{\xi}^{\pm} $ or   $\zeta^{\pm} < \tilde{\zeta}^{\pm} $\\
with $\tilde{\zeta}^{\pm}$ being the lowest value of $\zeta^{\pm}$ down to which the distributions of the light front variable, the polar angle, and the square of the transverse momentum of the particle under consideration can be fitted with Eq.\eqref{ZetaInt}, Eq.\eqref{CosInt} and Eq.\eqref{PtSqInt} respectively, assuming the form of the energy distribution as in Eq.\eqref{eqboltz}. We follow the $\chi^2$ minimization method for the fitting as incorporated in the ROOT (version 6.18/04) software package \cite{BRUN199781}. A successful fit would mean that the following relation \begin{equation}\label{ChiSq} 
\frac{\chi^2}{n.d.f} \sim 1.0
\end{equation}
holds for the three fits, where $n.d.f$ is the number of degree of freedom. 
\begin{figure*}[hbt!] 
\begin{subfigure}{0.5\textwidth}
\includegraphics[width=\linewidth]{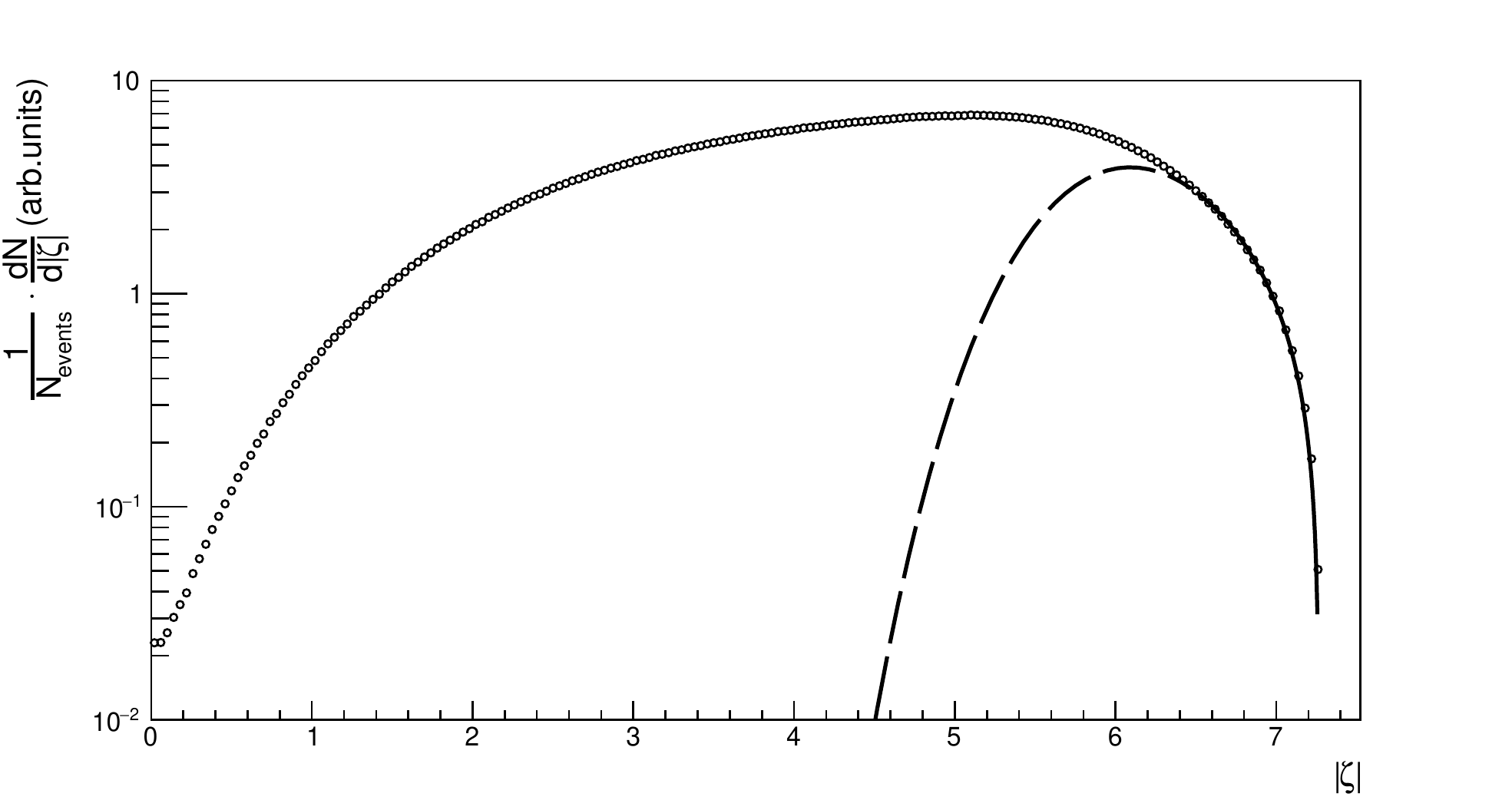}
\caption{$|\zeta^{\pm}|$ distribution of $\pi^{\pm}$; $\tilde{\zeta}^{\pm} =  6.50$ }\label{ZetaPion}
\end{subfigure}\hspace*{\fill}
\begin{subfigure}{0.5\textwidth}
\includegraphics[width=\linewidth]{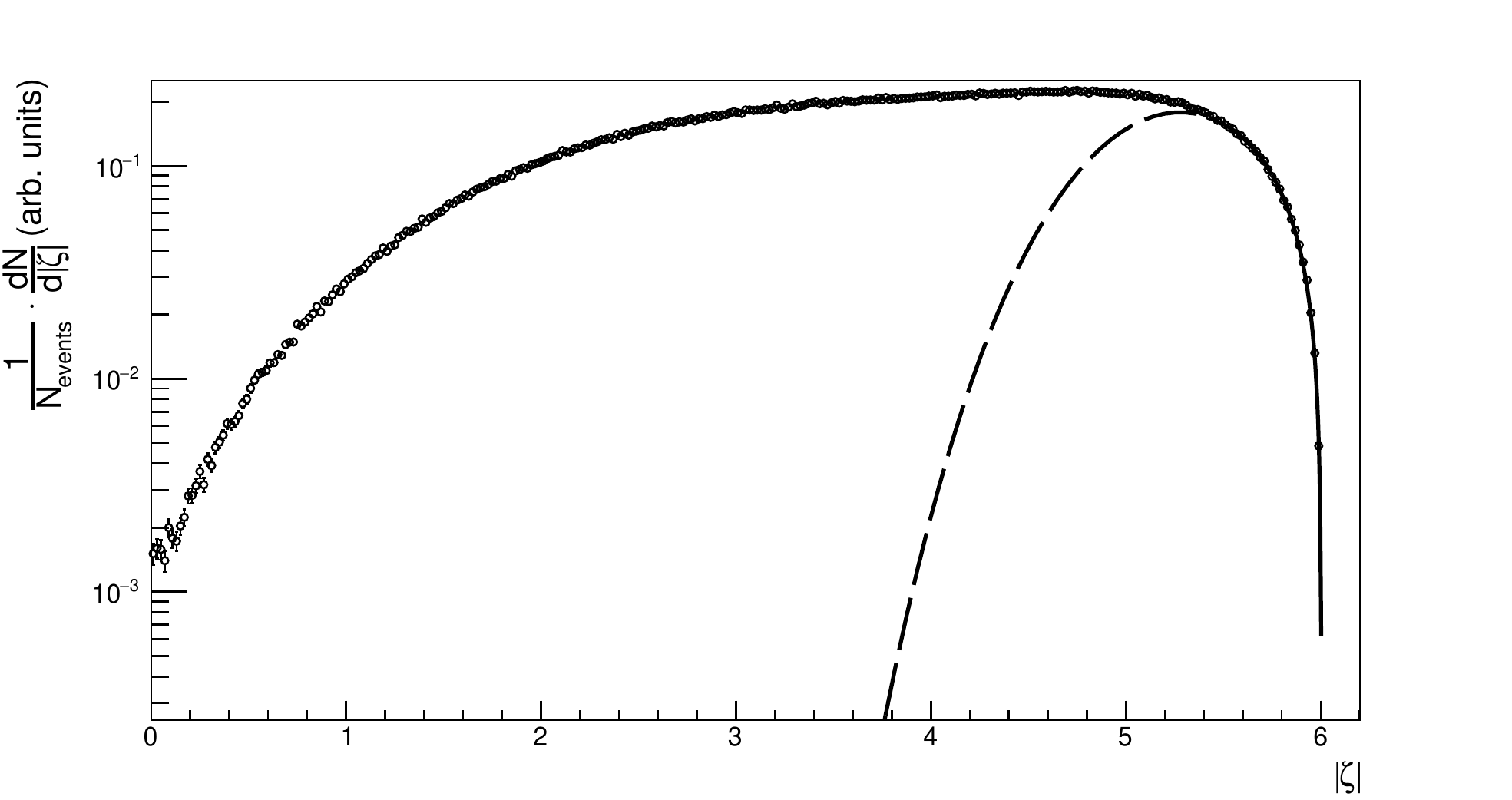}
\caption{$|\zeta^{\pm}|$ distribution of $K^{\pm}$; $\tilde{\zeta}^{\pm} =  5.50$}
\label{ZetaKaon}
\end{subfigure}
\medskip
\begin{subfigure}{0.5\textwidth}
\includegraphics[width=\linewidth]{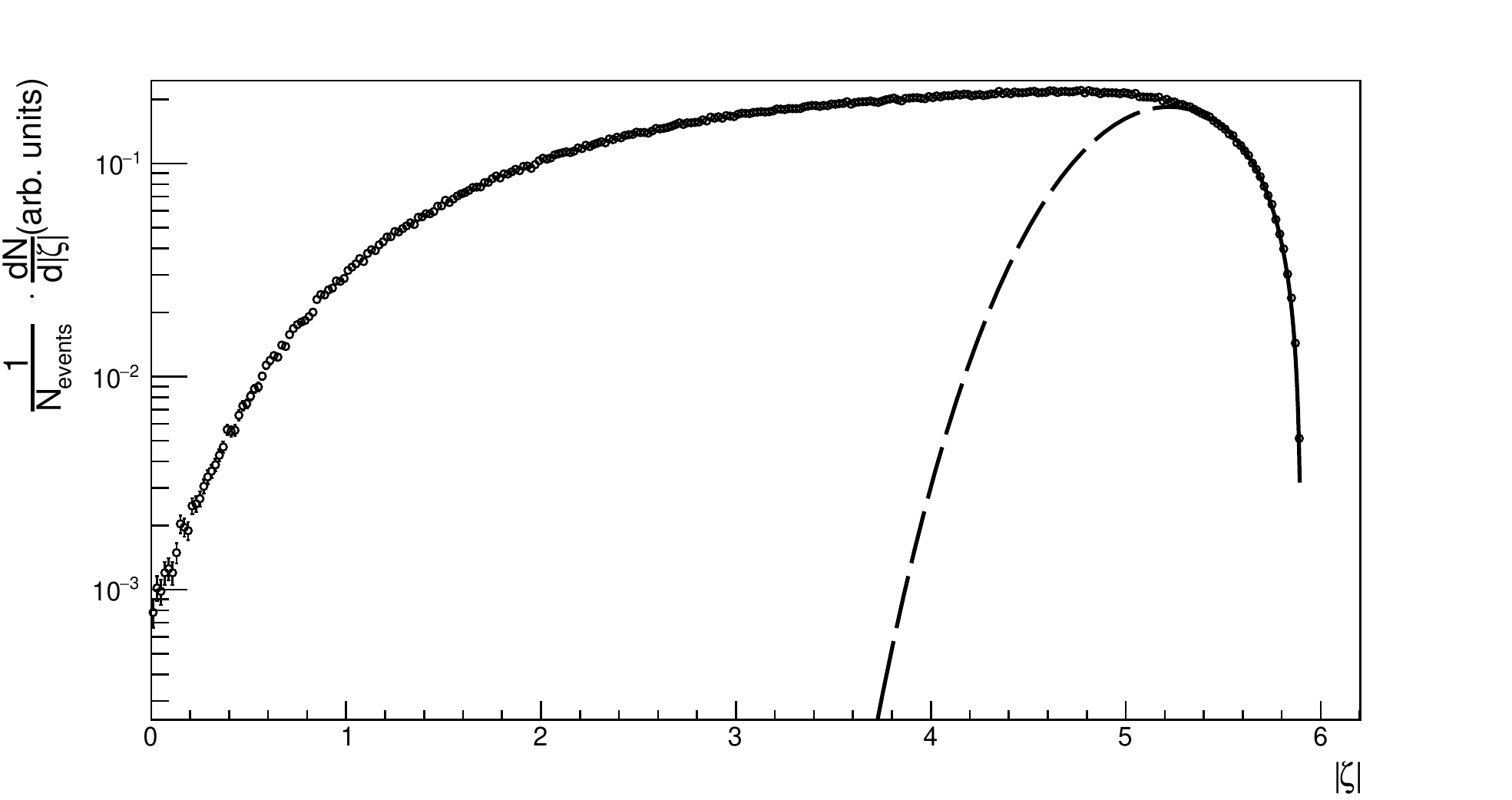}
\caption{$|\zeta^{\pm}|$ distribution of $\eta^{0}$; $\tilde{\zeta}^{\pm} =  5.40$}
\label{ZetaEta}
\end{subfigure}\hspace*{\fill}
\begin{subfigure}{0.5\textwidth}
\includegraphics[width=\linewidth]{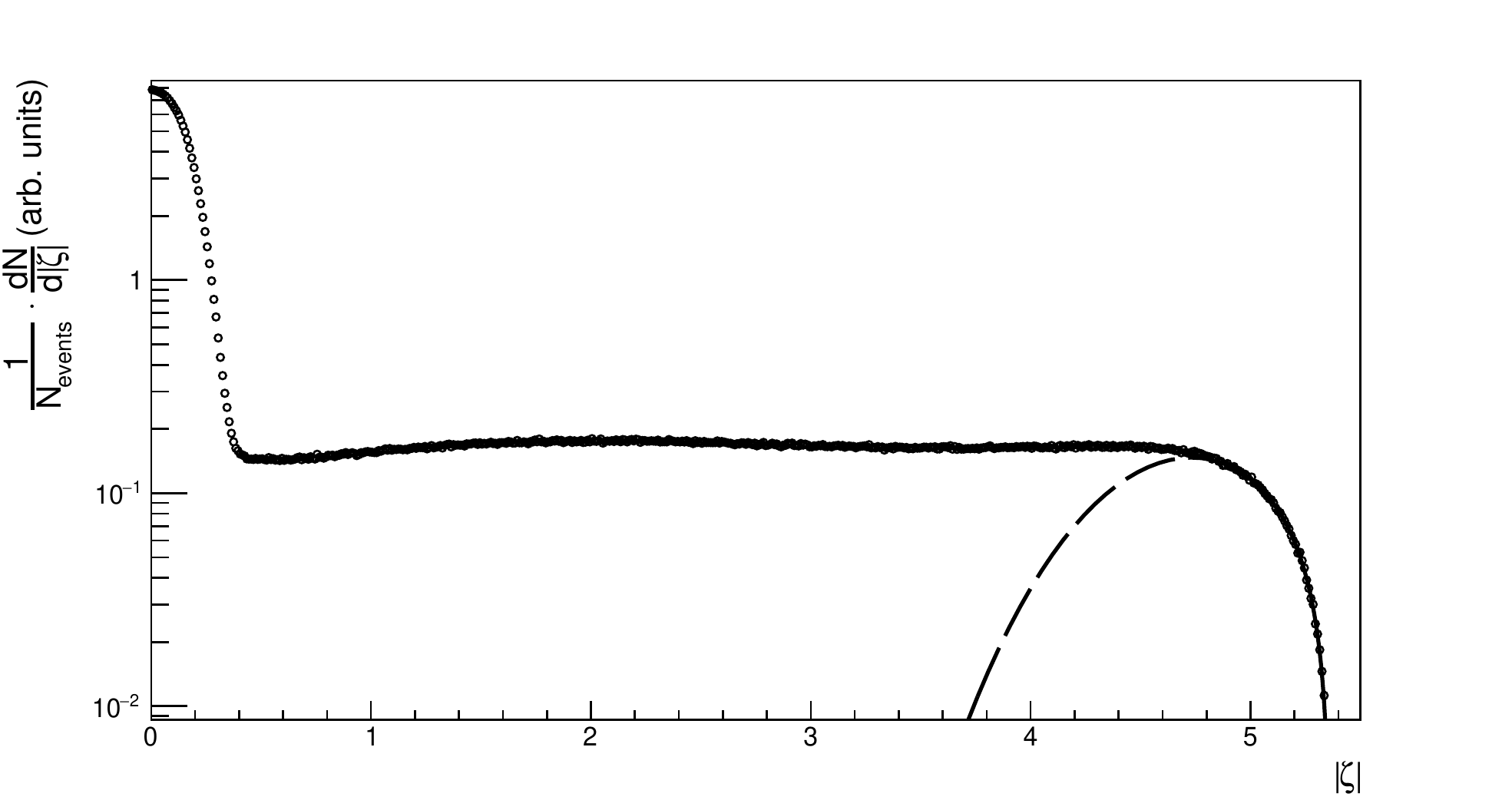}
\caption{$|\zeta^{\pm}|$ distribution of $p\bar{p}$; $\tilde{\zeta}^{\pm} =  4.85$}
\label{ZetaProton}
\end{subfigure}
\medskip
\begin{subfigure}[ht!]{0.5\textwidth}
\includegraphics[width=\linewidth]{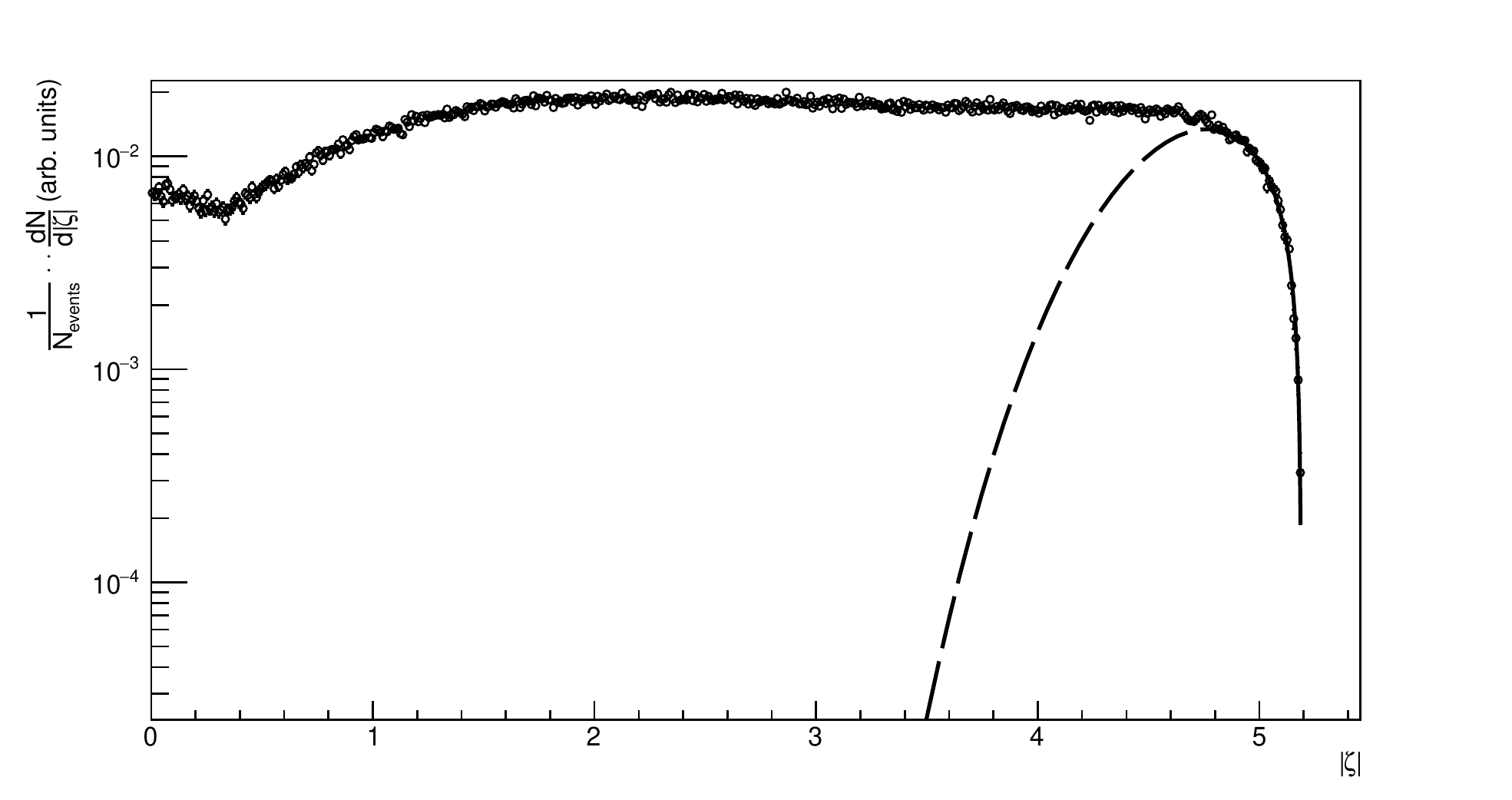}
\caption{$|\zeta^{\pm}|$ distribution of $\Lambda^{0}$; $\tilde{\zeta}^{\pm} =  4.90$}
\label{ZetaLambda}
\end{subfigure}\hspace*{\fill}
\begin{subfigure}{0.5\textwidth}
\includegraphics[width=\linewidth]{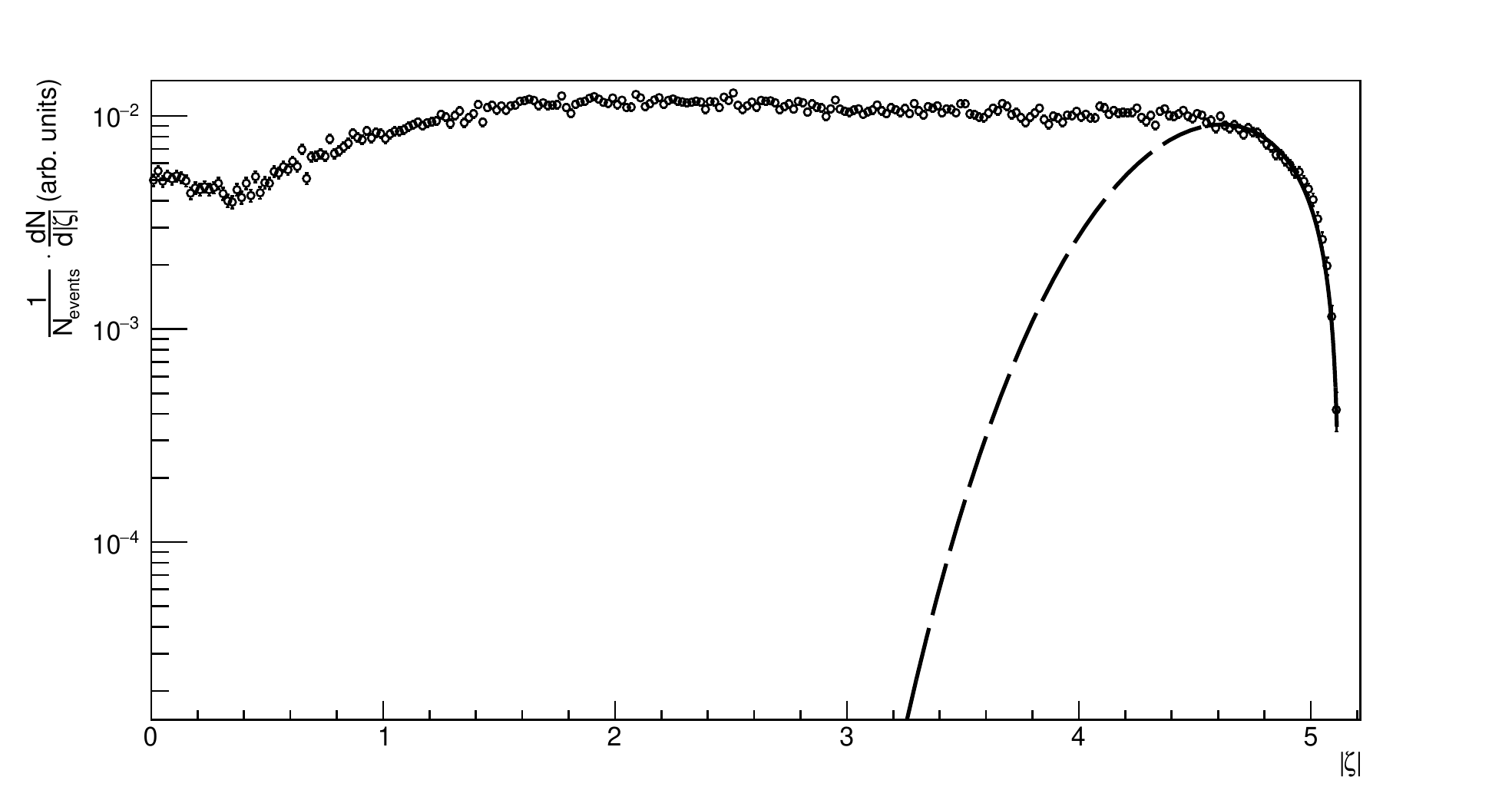}
\caption{$|\zeta^{\pm}|$ distribution of $\Sigma^{0}$; $\tilde{\zeta}^{\pm} =  4.55$}
\label{ZetaSigma}
\end{subfigure}
\caption{$|\zeta^{\pm}|$ distribution of particles fitted with Eq.\eqref{ZetaInt}. The solid curve is the result of the fit up to $\zeta_c$ and the dashed curve is the extrapolation of the fit beyond the corresponding $\zeta_c$ for each species of particles.} \label{UrQMDZetaPlots}
\end{figure*}
\begin{figure*}[hbt!] 
\begin{subfigure}{0.5\textwidth}
\includegraphics[width=\textwidth]{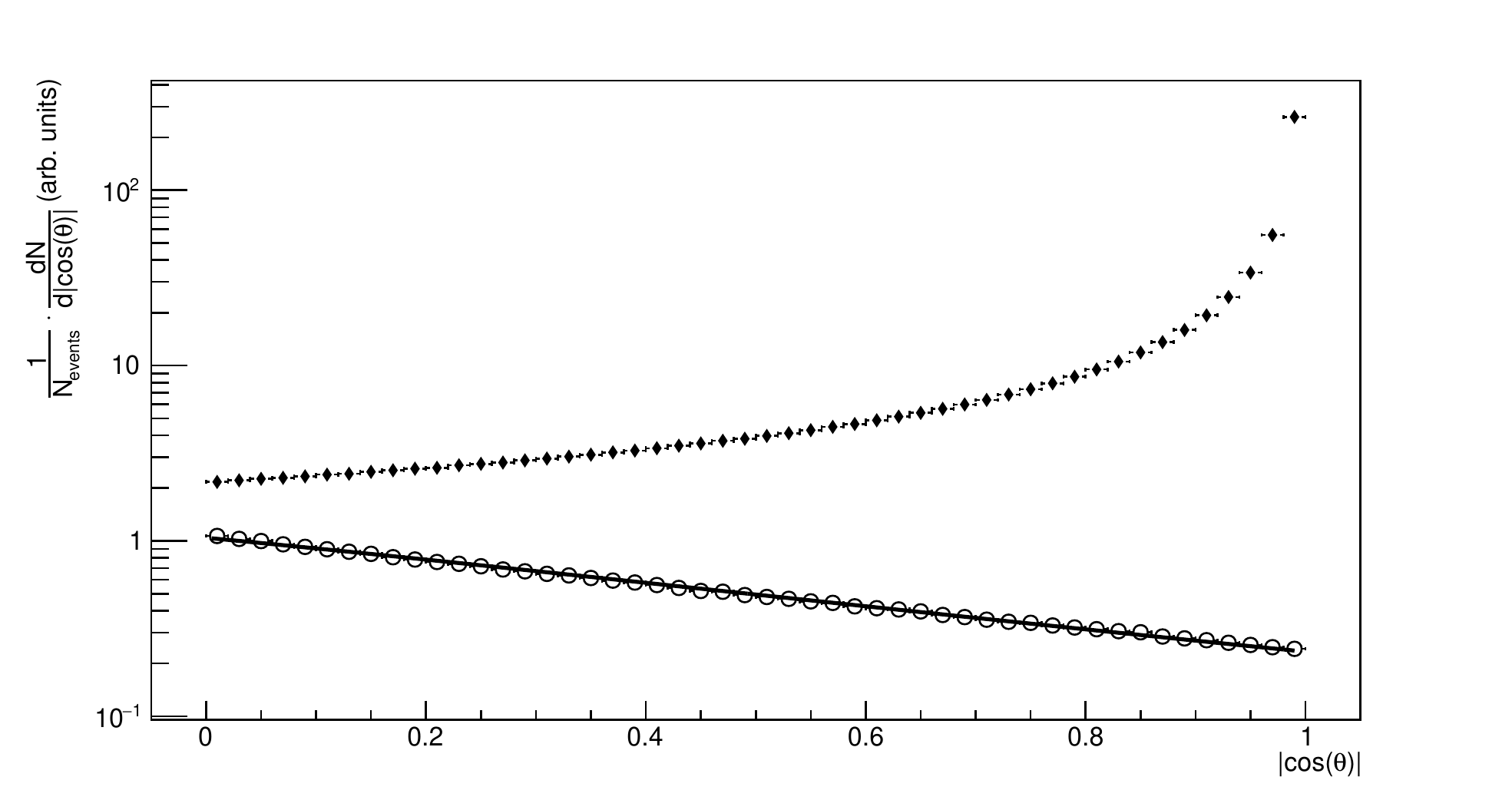}
\caption{ $|cos(\theta)|$ distribution of $\pi^{\pm}$; $\tilde{\zeta}^{\pm} 6.50$}\label{CosPion}
\end{subfigure}\hspace*{\fill}
\begin{subfigure}{0.5\textwidth}
\includegraphics[width=\linewidth]{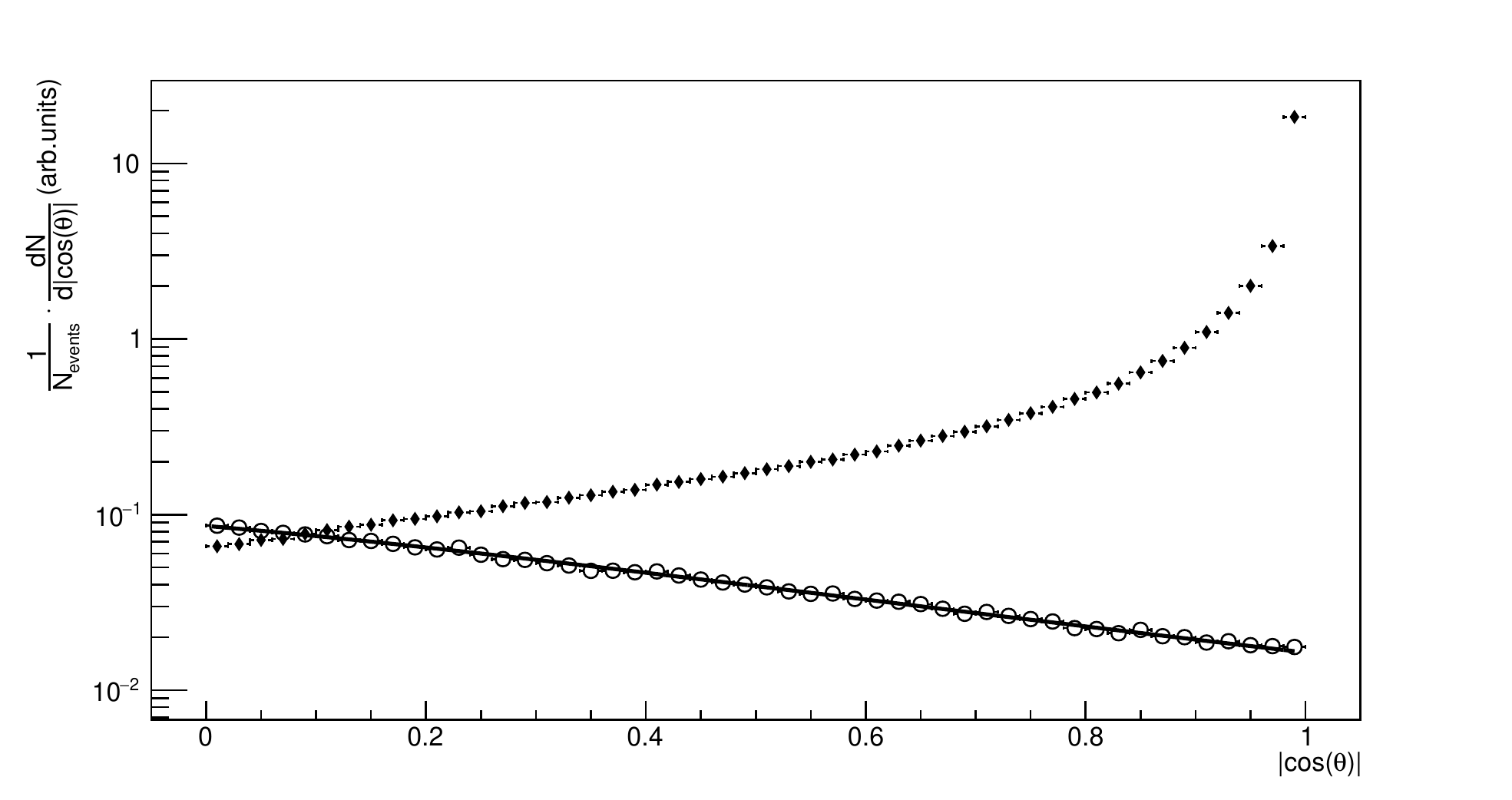}
\caption{$|cos(\theta)|$ distribution of $K^{\pm}$; $\tilde{\zeta}^{\pm} =  5.50$}
\label{CosKaon}
\end{subfigure}
\medskip
\begin{subfigure}{0.5\textwidth}
\includegraphics[width=\linewidth]{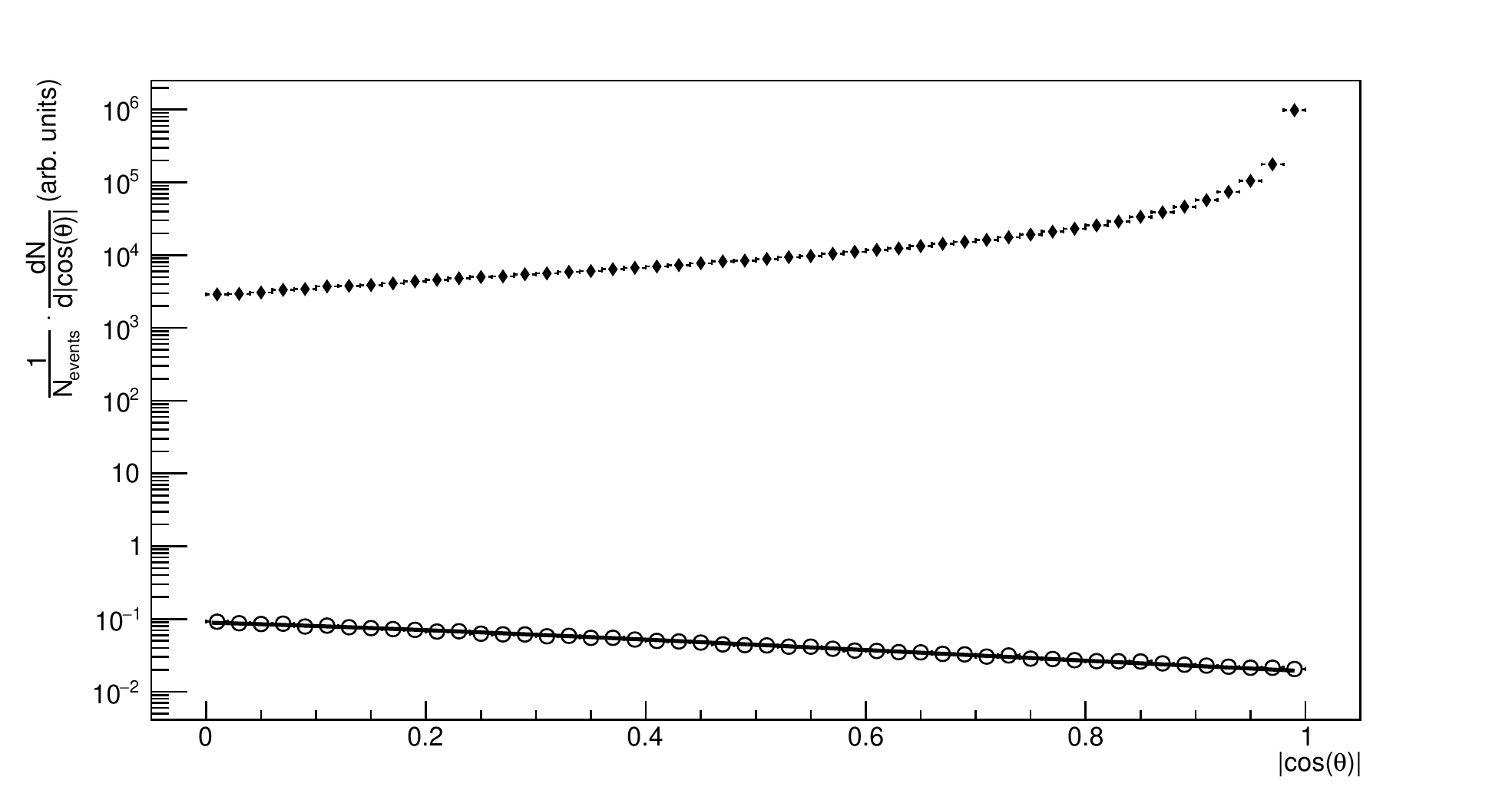}
\caption{$|cos(\theta)|$ distribution of $\eta^{0}$; $\tilde{\zeta}^{\pm} =  5.40$}
\label{CosEta}
\end{subfigure}\hspace*{\fill}
\begin{subfigure}{0.5\textwidth}
\includegraphics[width=\linewidth]{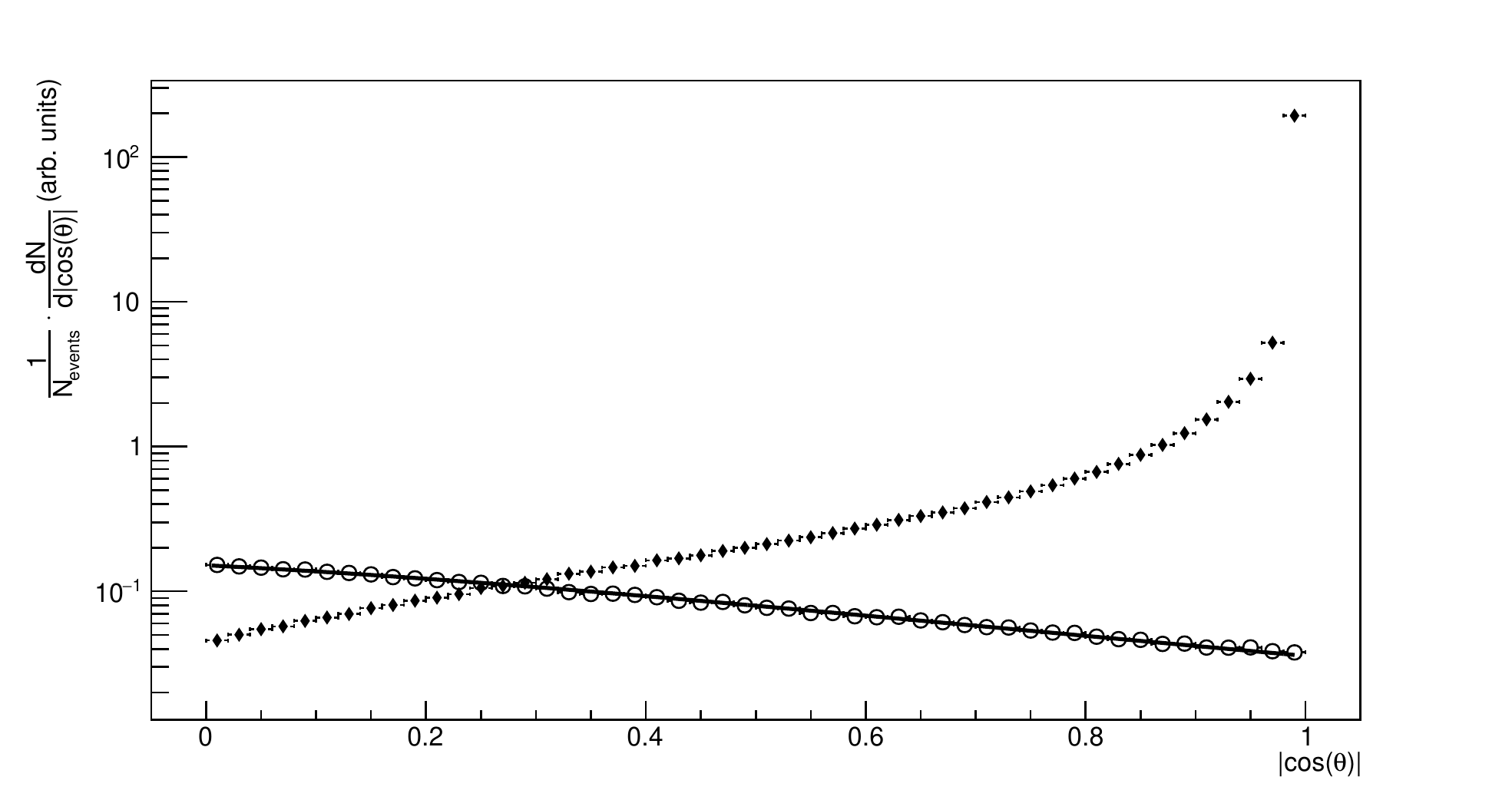}
\caption{$|cos(\theta)|$ distribution of $p\bar{p}$; $\tilde{\zeta}^{\pm} =  4.85$}
\label{CosProton}
\end{subfigure}
\medskip
\begin{subfigure}[ht!]{0.5\textwidth}
\includegraphics[width=\linewidth]{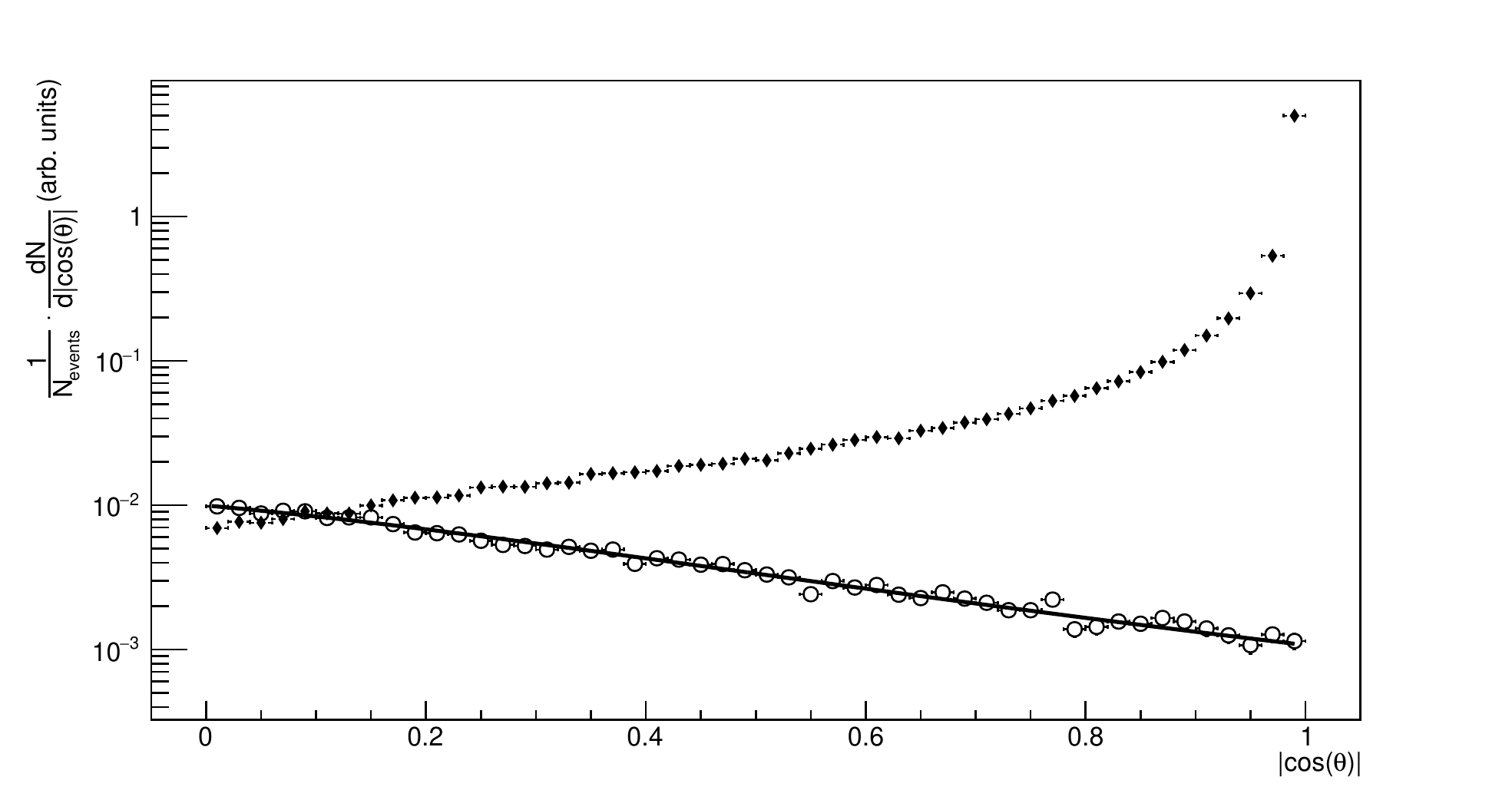}
\caption{$|cos(\theta)|$ distribution of $\Lambda^{0}$; $\tilde{\zeta}^{\pm} = 4.90$}
\label{CosLambda}
\end{subfigure}\hspace*{\fill}
\begin{subfigure}{0.5\textwidth}
\includegraphics[width=\linewidth]{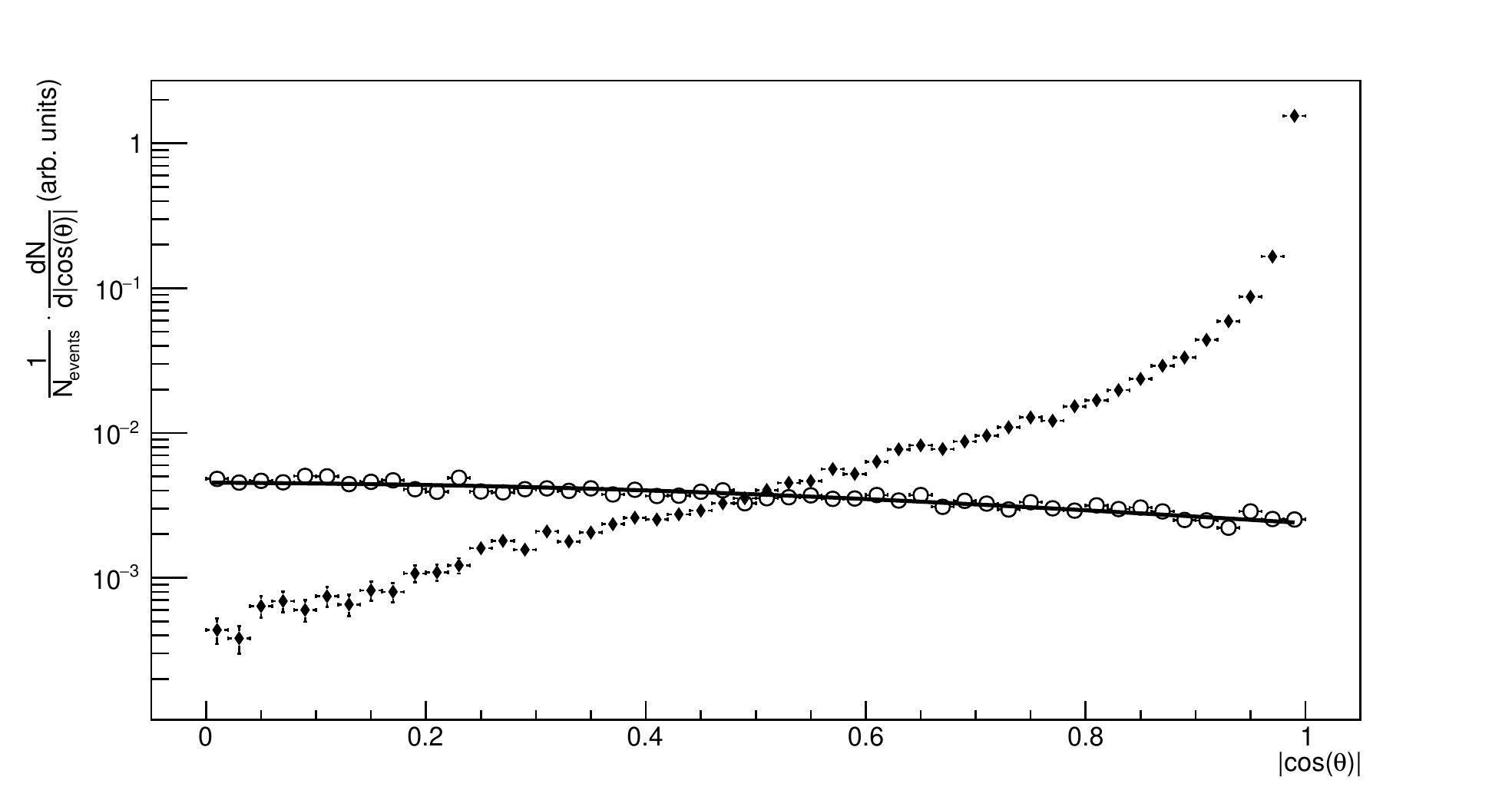}
\caption{$|cos(\theta)|$ distribution of $\Sigma^{0}$; $\tilde{\zeta}^{\pm} =  4.55$}
\label{CosSigma}
\end{subfigure}
\caption{Marked in open circles are the $|cos(\theta)|$ distribution of particles in region-1 fitted with Eq.\eqref{CosInt}. The solid curve is the result of the fit. The corresponding distribution of particles in Region-2 are marked with solid diamonds.} 
\label{UrQMDCosPlots}
\end{figure*}
Since the $\zeta^{\pm}$ and $cos(\theta)$ distributions are longitudinally symmetric in the system we are considering, we take the absolute value of $\zeta^{\pm}$ and $cos(\theta)$ to make our calculations in the positive hemisphere. The results we obtain will hold for both the hemispheres. 
\begin{figure*}[hbt!] 
\begin{subfigure}{0.5\textwidth}
\includegraphics[width=\textwidth]{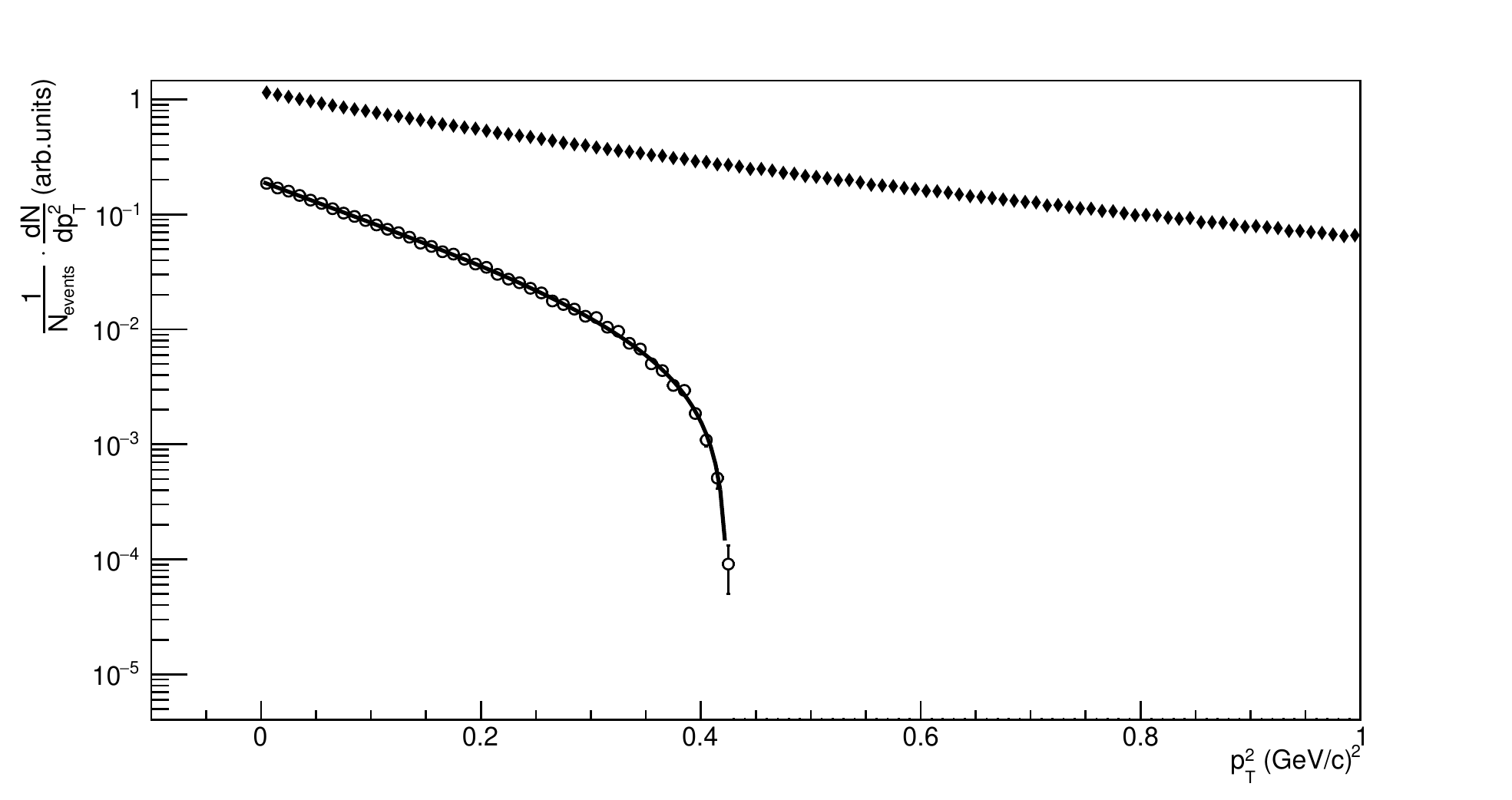}
\caption{$p_{T}^{2}$ distributions of $\pi^{\pm}$; $\tilde{\zeta}^{\pm} =  6.50$}
\label{PtSqPion}
\end{subfigure}\hspace*{\fill}
\begin{subfigure}{0.5\textwidth}
\includegraphics[width=\linewidth]{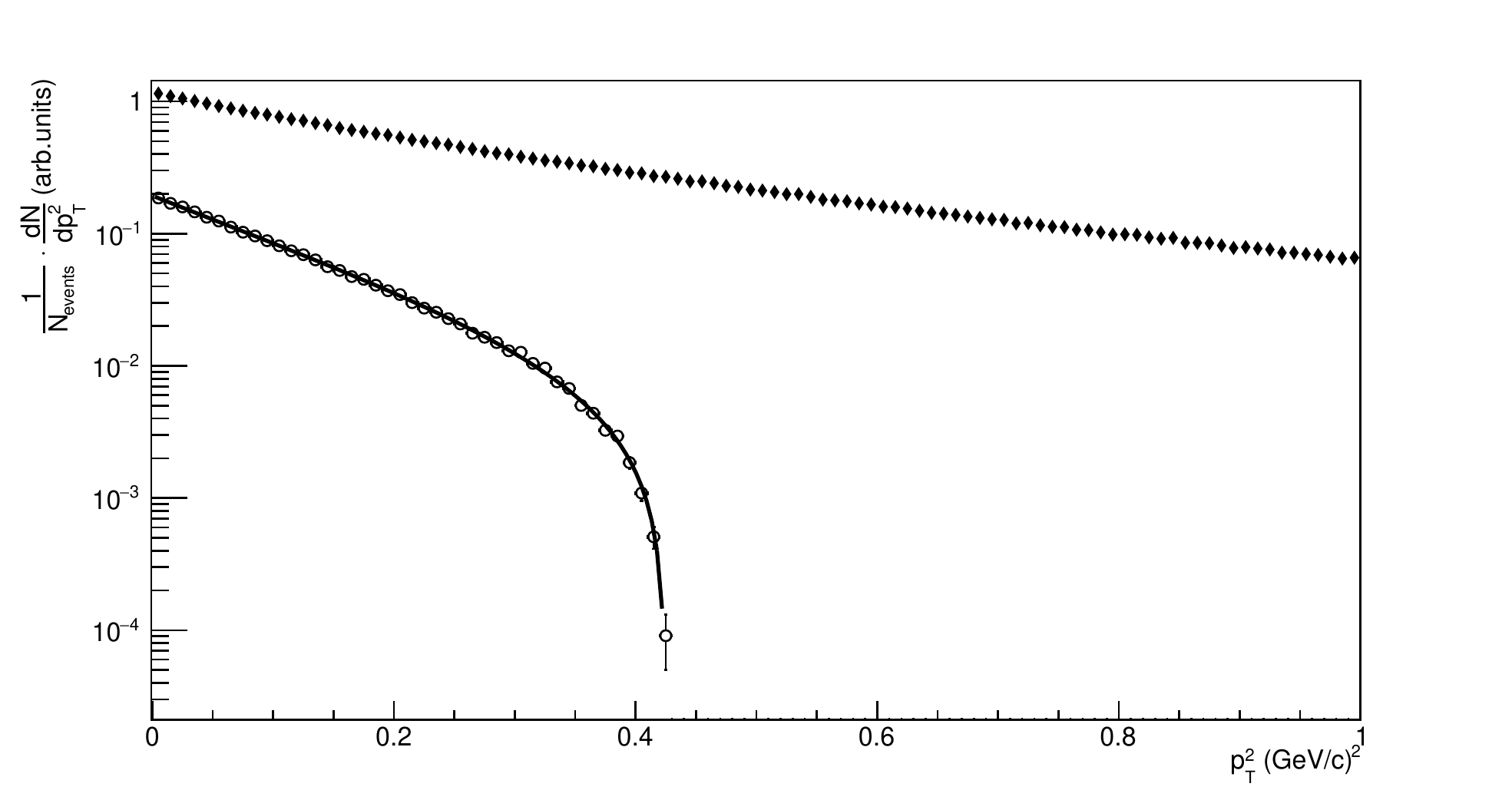}
\caption{$p_{T}^{2}$  distribution of $K^{\pm}$; $\tilde{\zeta}^{\pm} =  5.50$}
\label{PtSqKaon}
\end{subfigure}
\medskip
\begin{subfigure}{0.5\textwidth}
\includegraphics[width=\linewidth]{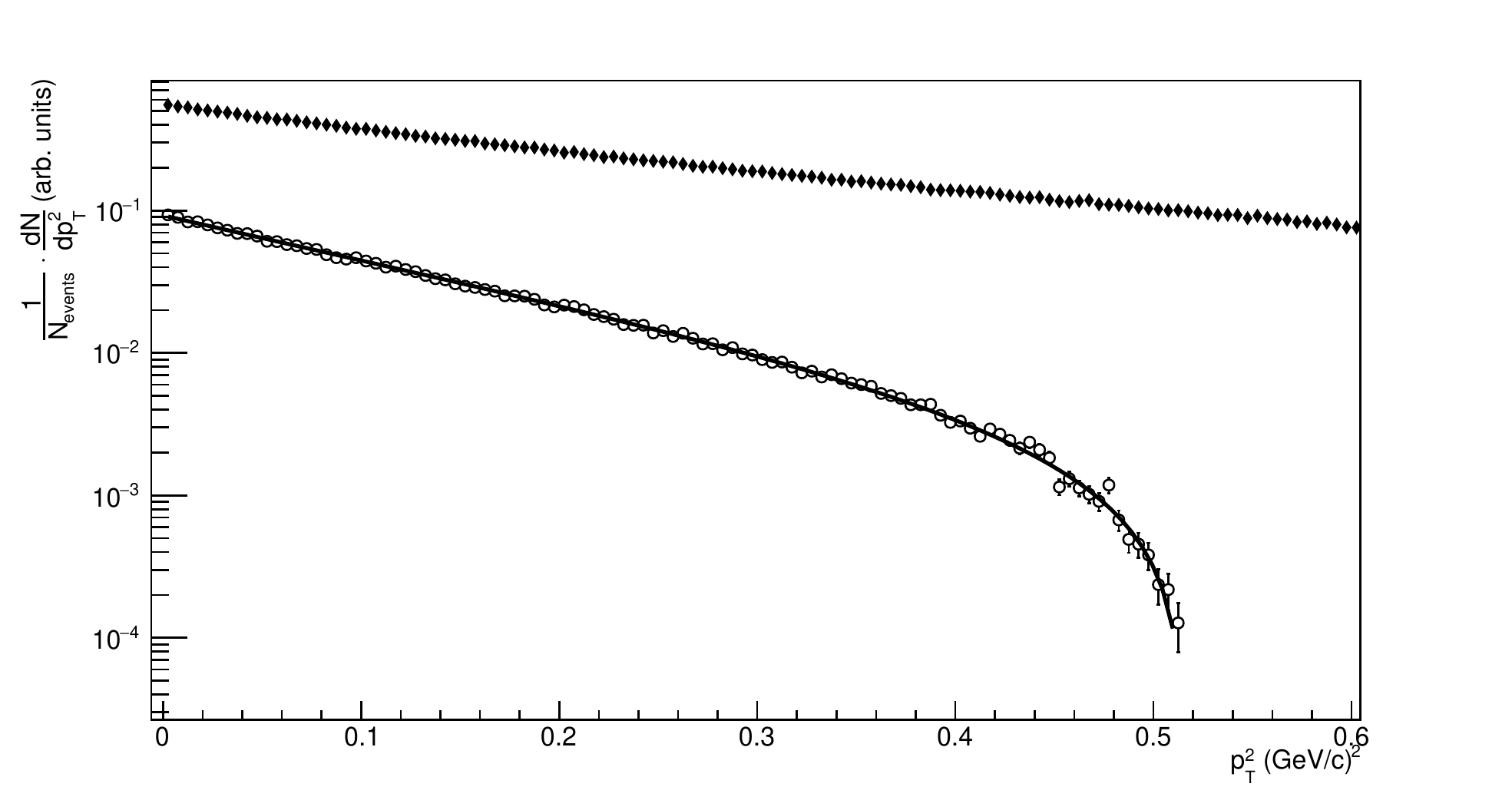}
\caption{$p_{T}^{2}$ distribution of $\eta^{0}$; $\tilde{\zeta}^{\pm} =  5.40$}
\label{PtSqEta}
\end{subfigure}\hspace*{\fill}
\begin{subfigure}{0.5\textwidth}
\includegraphics[width=\linewidth]{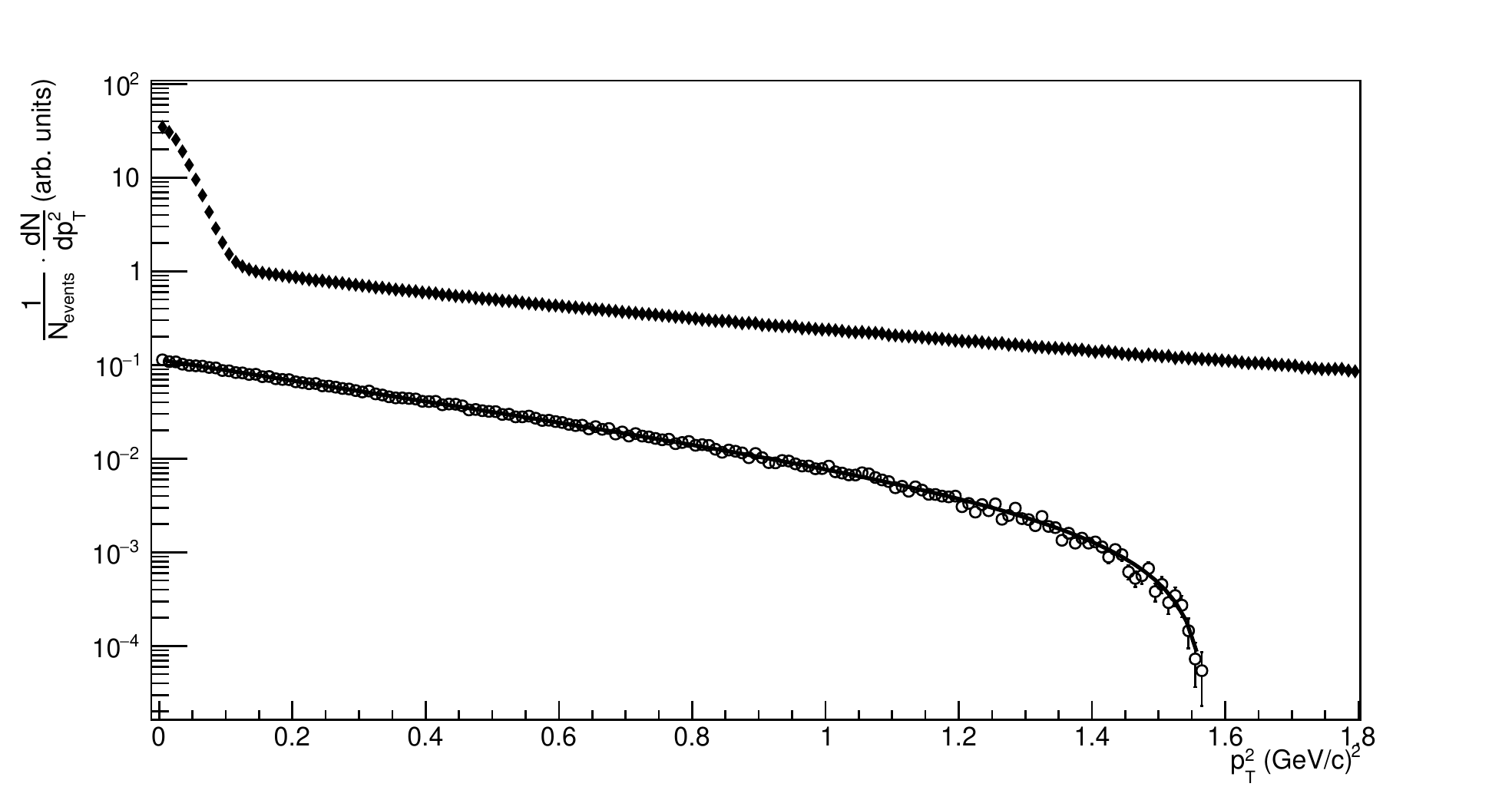}
\caption{$p_{T}^{2}$ distribution of $p\bar{p}$; $\tilde{\zeta}^{\pm} =  4.85$}
\label{PtSqProton}
\end{subfigure}
\medskip
\begin{subfigure}[ht!]{0.5\textwidth}
\includegraphics[width=\linewidth]{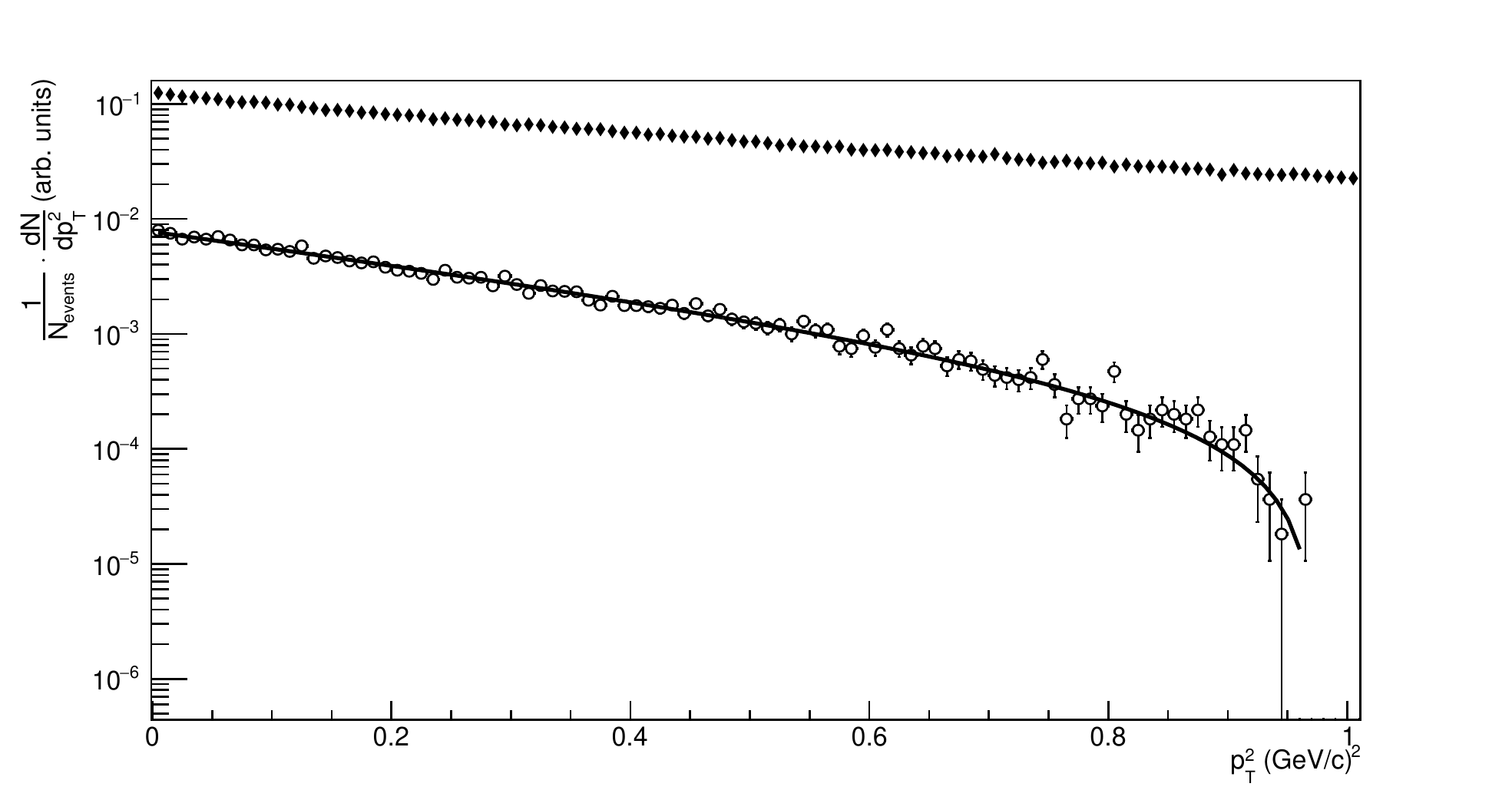}
\caption{$p_{T}^{2}$ distribution of $\Lambda^{0}$; $\tilde{\zeta}^{\pm} =  4.90$}
\label{PtSqLambda}
\end{subfigure}\hspace*{\fill}
\begin{subfigure}{0.5\textwidth}
\includegraphics[width=\linewidth]{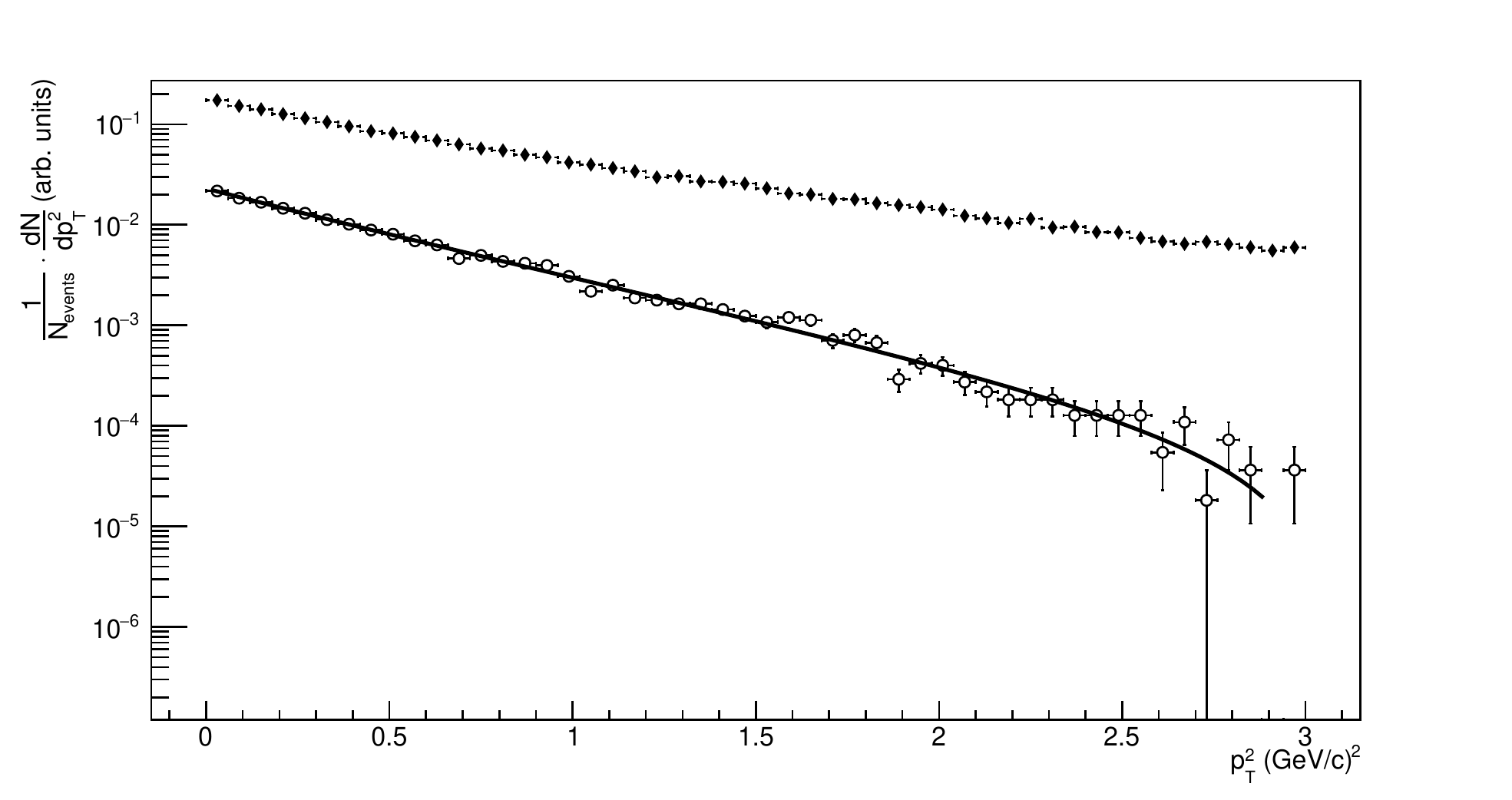}
\caption{$p_{T}^{2}$ distribution of $\Sigma^{0}$; $\tilde{\zeta}^{\pm} =  4.55$}
\label{PtSqSigma}
\end{subfigure}
\caption{Marked in open circles are the $p_{T}^{2}$ distribution of particles fitted with Eq.\eqref{PtSqInt} in region 1. The solid curve is the result of the fit. The distributions marked in solid diamonds are for the particles in Region-2.} \label{UrQMDPtSqPlots}
\end{figure*}
Exactly 55000 Au-Au collisions at $\sqrt{s} = 200$ GeV were simulated using the UrQMD  event generator (Version- 3.4). The inclusive single-particle distribution of the light front variable $|\zeta^{\pm}|$ are made for several species of particles. The $|\zeta^{\pm}|$ distributions of the UrQMD hadrons are then fitted with Eq.\eqref{ZetaInt}. The lowest value of $|\zeta^{\pm}|$ to which this could be achieved successfully, is taken as $|\tilde{\zeta}^{\pm}|$. Based on this $|\tilde{\zeta}^{\pm}|$, we divide the phase space of the corresponding particle into Region-1 and Region-2 as we defined earlier. The next step is to make $|cos(\theta)|$ and $p_T^2$ distributions for the two groups. These two distributions of the particles in Region-1 are fitted with Eq.\eqref{CosInt} and Eq.\eqref{PtSqInt} respectively. If the fit can be done successfully for the three distributions, the corresponding value of $\tilde{\zeta}$ is taken as the constant value $\zeta_c$ of the light front variable for the specific particle specie. The set of particles inside the paraboloid defined by $\zeta_c$ are then considered as a thermalised group. If any one of the three fits is not successful, the procedure is repeated with a larger value of $\tilde\zeta$ either until a $\zeta_c$ is found or the range of $\zeta$ is exhausted. We say that the light front analysis fails to find a constant light front surface in the latter case for the system under consideration.
\section{Results}\label{sec:3} 
The scheme of analysis described in the previous section was performed for $\pi^{\pm}$ mesons from the UrQMD simulated events. It was observed that if we perform the fit up to $\zeta_c = 6.50$ and divide the whole of the charged pions in the phase space in to two groups with $|\zeta^{\pm}| > 6.50 $ and  $|\zeta^{\pm}| < 6.50 $, we could fit the $|cos(\theta)|$ and $p_{T}^{2}$ distributions of the $\pi^{\pm}$ mesons in Region-1 with Eq.\eqref{CosInt} and Eq.\eqref{PtSqInt}. The results of the fits are shown in Fig.\ref{ZetaPion}, Fig.\ref{CosPion} and Fig.\ref{PtSqPion} together with the corresponding distributions in Region-2 for the last two cases. Hence the surface of $\zeta_c  = 6.50$ in the phase space of $\pi^{\pm}$ in Au-Au collisions simulated using UrQMD at $\sqrt{s} = 200$ GeV separates out a group of charged pions that has reached thermal equilibrium. To see the effect of the mass of the particle on the nature of the proposed light front surface, we performed a similar analysis for five other species of hadrons ($K^{\pm}$, $\eta^0$, $p(\bar{p})$, $\Lambda^0$ and $\Sigma^{0}$) from the same set of collisions. We could always find a $\zeta_c$ for all the considered species of UrQMD hadrons. The $|\zeta^{\pm}|$  distributions of the particles with the fit using the Eq.\eqref{ZetaInt} up to the estimated value of  $\zeta_c$ are shown in Fig.\ref{UrQMDZetaPlots}. The corresponding $|cos(\theta)|$ distributions with the fit using the Eq.\eqref{ZetaInt} and $p_T^2$ distributions with the fit using the Eq.\eqref{PtSqInt} are shown respectively in Fig.\ref{UrQMDCosPlots} and Fig.\ref{UrQMDPtSqPlots}. Note that the uncertainties in the distributions are statistical. The $|cos(\theta)|$ distributions in Region-1 is isotropic with respect to that in Region-2 and the slopes of the $p_T^2$ distributions are different for two regions as well for all the species we considered. Thus a comparison of the shape of the $cos(\theta)$ and $p_T^2$ distributions in the two regions together with the successful description of the distributions in Region-1 using the statistical model confirms that the observation made for the charged pions do stay for particles of heavier masses as well. following the thermal Boltzmann distribution which suggests that the charmed mesons also got thermalised. A summary of the extracted temperatures is reported in Table \ref{UrQMDTempratureTable}
\begin{table*}[hbt!]
\begin{center}
\begin{tabular} {llllllll}
\hline\hline \noalign{\smallskip}
Species       & $\zeta_{c}$ & $T_{\zeta}$ (MeV) & $\chi^2/n.d.f$&  $T_{p_T^2}$ (MeV)& $\chi^2/n.d.f$  & $T_{cos(\theta)}$ (MeV) & $\chi^2/n.d.f$\\
\noalign{\smallskip}\hline\noalign{\smallskip}
$\pi^{\pm}$   & 6.50 &  96$\pm$4 &2/18 &  87 $\pm$1 & 43/68  &  72$\pm$1 &53/48\\
$K^{\pm}$     & 5.50 & 183$\pm$6 &3/23 & 147 $\pm$3 & 7/40   & 129$\pm$2 &36/48\\
$\eta^{0}$    & 5.40 & 187$\pm$5 &3/23 & 147 $\pm$1 & 70/102 & 123$\pm$2 &49/48\\
$p(\bar{p})$  & 4.85 & 295$\pm$3 &24/50& 235 $\pm$1 & 155/155& 200$\pm$2 &38/48\\
$\Lambda^{0}$ & 4.90 & 221$\pm$14&16/27& 177 $\pm$4 & 95/95  & 161$\pm$6 &45/48\\
$\Sigma^{0}$  & 4.55 & 291$\pm$9 &23/27& 199 $\pm$4 & 40/46  & 142$\pm$5 &47/48\\
\noalign{\smallskip}\hline\hline
\end{tabular}
\end{center}
\caption{Temperatures obtained from the light front analysis of UrQMD hadrons.}
\label{UrQMDTempratureTable}
\end{table*}
\section{Conclusions and discussions}\label{sec:4}
The main aim of this analysis was to demonstrate that the light front analysis scheme as proposed in \cite{Garsevanishvili78, Garsevanishvili79} and performed in \cite{Amaglobeli99, Djobava03} etc gives qualitatively similar results at RHIC energies and for particles heavier than pions within the UrQMD framework. It is concluded from the results that the light front analysis scheme works for particles heavier than pions in the heavy-ion collisions at RHIC energy within the UrQMD model. The vanishing of the relative isotropy in the polar angle distribution of particles with $\zeta > \zeta_c$ for heavier species than pions as argued in \cite{Levchenko} is discarded. A thermalisation is observed to have reached in the collisions considered in the analysis and the temperatures extracted with it varies with respect to the mass of the particle. Hence it is clear that there exists a nontrivial integral relationship between the longitudinal and transverse kinematic variables of the inclusively produced particles in relativistic heavy-ion collisions. A criterion of convergence of the temperatures obtained from the $\zeta$, $cos(\theta)$ and $p_T^2$ distributions may be implemented while performing the analysis with the experimental data. However such a convergence is not observed for the earlier experimental analysis at low energies. As mentioned in \cite{Chkhaidze2006}, the ALEPH collaboration observed the maxima in $\xi$ distributions with  $\xi =\ln(p/p_{max})$  of secondary hadrons in $e^{+}e^{-}$ collisions, which coincide to high precision with predictions of the perturbative QCD (See Section 4.1.1 in \cite{BARATE19981}). The accuracy of coincidence increases when next to leading order corrections are taken into account. The conclusion was that the shapes of these $\xi$ distributions are related to the details of the underlying dynamics. A similar test of the light front $\zeta$ distributions might be performed to quantify up to what level the shape of the $\zeta$ distributions reflects the dynamics of in the collision process at RHIC and LHC. From the phenomenological studies, it is known that the charged hadrons at sufficiently low transverse momentum follow approximately thermal momentum distributions. This is reproduced by blast wave models (\cite{PhysRevC.50.1675, PhysRevC.48.2462} \cite{PhysRevLett.92.112301, PhysRevC.79.034909}) and hydrodynamical simulations of heavy-ion collisions \cite{doi:10.1063/1.1843595, HEINZ2002269}. From this perspective, the finding with the light front analysis is not a surprise and it endorses the already known understanding of thermalisation. However, one can easily extract the temperatures directly using Boltzmann statistics from this scheme of analysis with three specific inclusive single-particle distributions. The values of temperatures extracted from the light front analysis may be taken as a pointer to the phase transition and deconfinement effects. Those temperatures obtained from the UrQMD based light front analysis hints towards the possibility of such a deconfined state of quarks and gluons at RHIC energies. Further studies with experimental and simulated data are required to conclude the relationship between the formation of a deconfined state of quark-gluon plasma in the collisions and the results of the light front analysis. Additionally, the study of two and multi-particle correlations of particles inside and outside the light front paraboloid might be a good tool to explore up to what extent the chosen surface is unique and 'critical' as mentioned in \cite{Garsevanishvili79}. With the current understanding, the criticality of the chosen light front paraboloidal surface is limited to the successful description of the distributions of particles it encapsulates, with Boltzmann statistics and a resulting isotropic angular distribution irrespective of the mass of the particle at RHIC energies. Note that the conclusions are drawn within the context of a phenomenological model. The implementation of the light front analysis using the experimental data from the heavy-ion collisions at RHIC and LHC is therefore compelling which requires the incorporation of kinematical and acceptance restrictions of the detectors in to the calculations. 
\section*{Acknowledgements}
I would like to thank Prof. Marcus Bleicher (UrQMD collaboration) for the consent to use the UrQMD source code. I thank Dr Bharati Naik (IIT Bombay), Prof. Kajari Mazumdar(TIFR, Mumbai), Dr Preeti Dhankher (University of California, Berkeley) and Dr Souvik Priyam Adhya (Charles University, Prague) for helping me with the computational resources.
\bibliography{article}
\end{document}